\documentclass[twocolumn,showpacs,preprintnumbers,amsmath,amssymb]{revtex4}
\usepackage{graphicx}% Include figure files
\usepackage{dcolumn}% Align table columns on decimal point
\usepackage{bm}% bold math

\begin{document}

\preprint{APS/123-QED}
\title{  Essence of inviscid shear instability: a point view of vortex dynamics}
\author{Liang Sun}
\email{sunl@ustc.edu.cn;sunl@ustc.edu} \affiliation{1. School
of Earth and Space Sciences, 2. Dept. of Modern Mechanics,\\
 University of Science and Technology of China, Hefei, 230027, China.}

\date{\today}
\begin{abstract}

The essence of shear instability is fully revealed both
mathematically and physically. A general sufficient and necessary
stable criterion is obtained analytically within linear context.
It is the analogue of Kelvin-Arnol'd theorem, i.e., the stable
flow minimizes the kinetic energy associated with vorticity. Then
the mechanism of shear instability is explored by combining the
mechanisms of both Kelvin-Helmholtz instability (K-H instability)
and resonance of waves. It requires both concentrated vortex and
resonant waves for the instability. The waves, which have same
phase speed with the concentrated vortex, have interactions with
the vortex to trigger the instability. We call this mechanism as
"concentrated vortex instability". The physical explanation of
shear instability is also sketched. Finally, some useful criteria
are derived from the theorem. These results would intrigue future
works to investigate the other hydrodynamic instabilities.

\end{abstract}
 \maketitle

The hydrodynamic instability  is a fundamental problem in many
fields, such as fluid dynamics, astrodynamics, oceanography,
meteorology, etc. There are many kinds of hydrodynamic
instabilities, e.g., shear instability due to velocity shear,
thermal instability due to heating, viscous instability due to
viscosity and centrifugal instability due to rotation, etc. Among
them, the shear instability is the most important and the simplest
one, which has been intensively explored (see
\cite{Drazin1981,Huerre1998,CriminaleBook2003} and references
therein). Both linear and nonlinear stabilities of shear flow have
been considered, and some important conclusions have also be
obtained from the investigations.

On the one hand, the nonlinear stability of shear flow has been
investigated via variational principles. Kelvin \cite{Kelvin1875}
and Arnol'd \cite{Arnold1965b,Arnold1969,ArnoldBook1998} have
developed variational principles for two-dimensional inviscid flow
\cite{Saffman1992}. They showed that the steady flows are the
stationary solutions of the energy $H$. And if the second
variation $\delta^2 H$ is definite, then the steady flow is
nonlinearly stable. Moreover, Arnol'd proved that the flow is
linearly stable provided that $\delta^2 H$ is positive definite,
and he also proved two nonlinear stability criteria
\cite{Arnold1969,Saffman1992,ArnoldBook1998,Vladimirov1999}.
However, $\delta^2 H$ is always indefinite in sign, except for two
special cases (see \cite{Vladimirov1999} and references therein).
Though variational principle is indeed powerful, it is also
inconvenient for real applications due to the lack of the explicit
expressions in both $H$ and the stability criteria.

On the other hand, the linear stability of shear flow has also
been investigated via Rayleigh's equation. Within the linear
context, there are three important general stability criteria,
which are Rayleigh's criterion \cite{Rayleigh1880}, Fj{\o}rtoft's
criterion \cite{Fjortoft1950} and Sun's criterion
\cite{SunL2006a}. As all the criteria have explicit expressions,
they are more convenient in real applications, and are widely used
in many fields. Based on the former investigations, Sun
\cite{SunL2006a} also pointed out that the flow is stable for
Rayleigh's quotient $I(f)>0$ (see
Eq.(\ref{Eq:stable_paralleflow_sun_Energy}) behind). However, the
sufficient criterion for instability still lacks.
% So the flow might be unstable only for $I(f)<0$.

To understand the shear instability, some mechanisms were
suggested. Among them, Kelvin-Helmholtz instability (K-H
instability) is always taken as a prototype, which is physically
explained as the instability of a sheet vortex
\cite{Batchelor1967}. An another mechanism of instability is due
to the resonance of waves
\cite{Craik1971,Butler1992,Baines1994,Staquet2002,CriminaleBook2003}.
Butler and Farrell \cite{Butler1992} clearly showed with numerical
simulations that the resonance introduces an algebraic growth term
into the temporal development of a disturbance. Baines and
Mitsudera \cite{Baines1994} also used broken-line profile velocity
as a prototype to explain the interaction of waves. Their
explanation is so brilliant that it can explain why the
instability occurs for a finite range of wavenumber and how the
waves amplify each other. However, both mechanisms are independent
of the basic flows, for that Kelvin-Helmholtz model deals only
with vortex and resonance mechanism only considers the waves. Thus
the relationships between those mechanisms and the shear flows are
still covered.

Overview the former investigations, the essence of the shear
instability is remained to be elucidated. Though $H$ is very
important in variational principles, both the explicit expression
and the meaning of it have not been revealed before. The
connection between linear and nonlinear stability criteria should
be retrieved explicitly. The relationships between instability
mechanisms and the physical explanation for shear instability are
needed. The aim of this letter is to fully reveal the essence of
the shear instability by investigating the inviscid shear flows in
a channel. And other instabilities in hydrodynamics may also be
understood via the investigation here.

\iffalse
\begin{figure}
%\includegraphics{fig1}% Here is how to import EPS art
  \includegraphics[width=6cm]{stable_parallelflow_sketch.eps}
\caption{ Sketch of parallel flow.}
\label{Fig:stable_sketch_vorticity_profile}
\end{figure}
%
\fi

\begin{figure}
  \includegraphics[width=6cm]{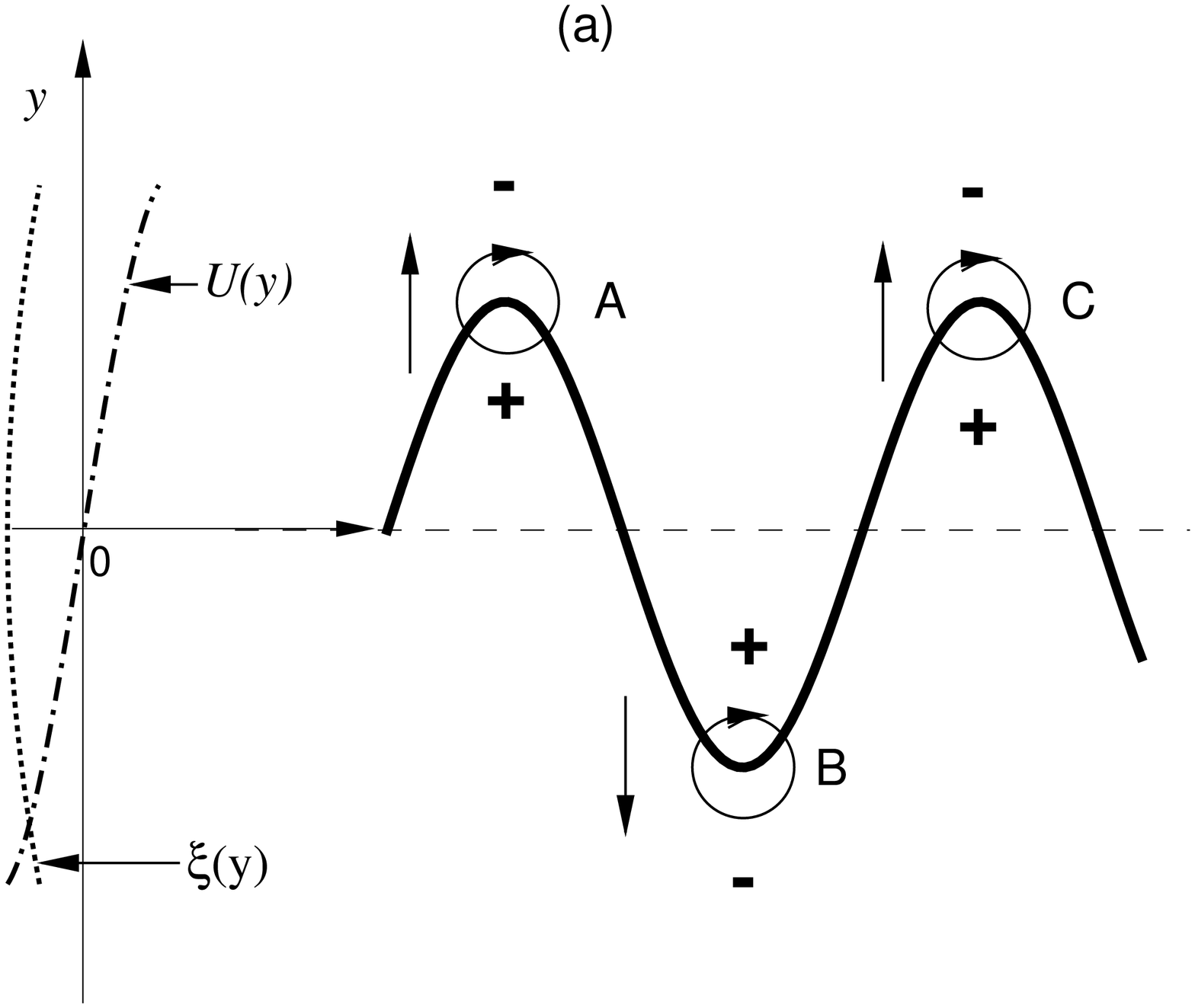}
  \includegraphics[width=6cm]{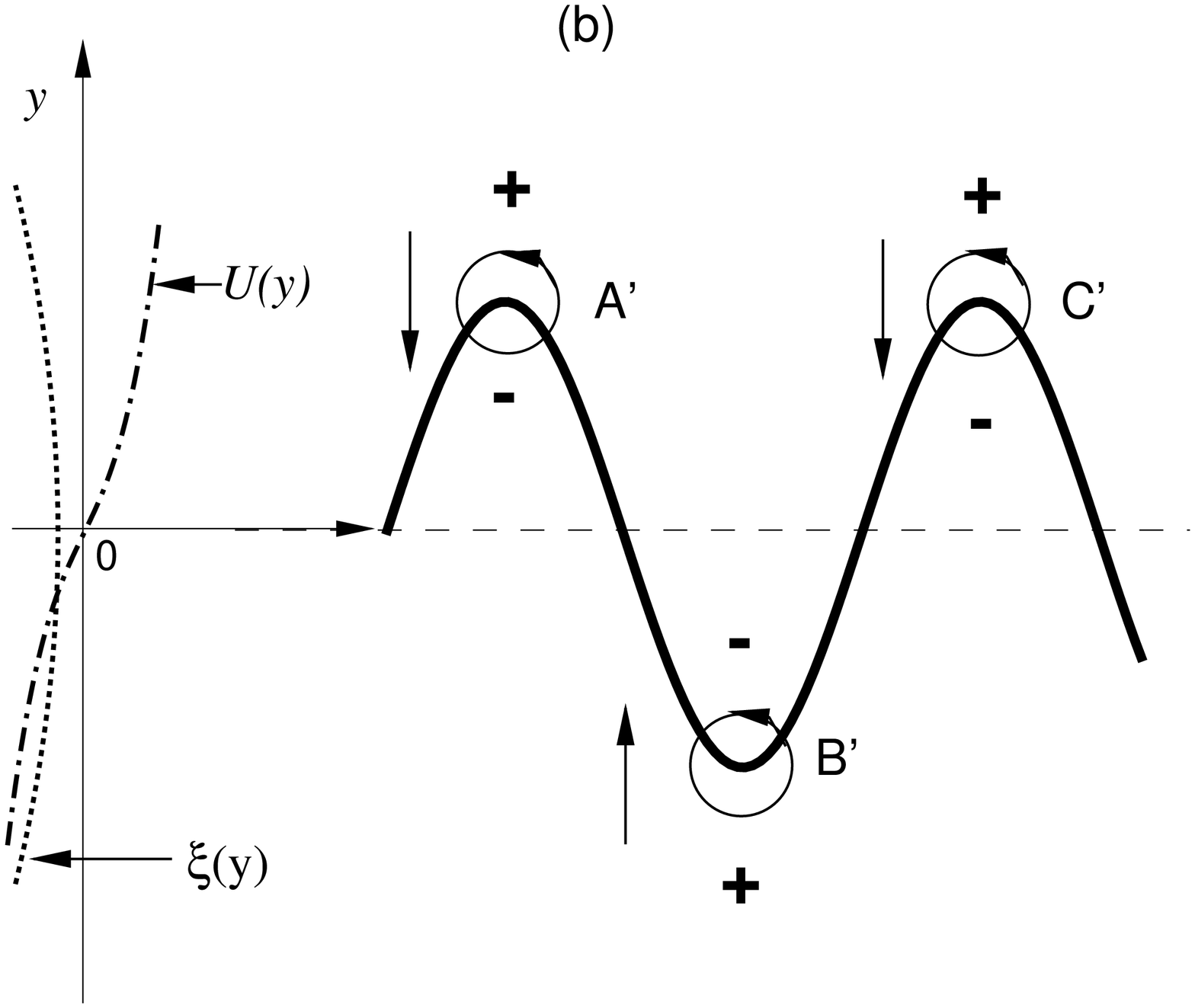}
\caption{Sketch of shear instability: physical interpretation.
Left parts depict the profiles of velocity $U(y)$ and vorticity
$\xi(y)$, right ones depict the disturbance of vorticities. The
unstable veloctiy profile $U(y)$ has a local maximum in vorticity
$\xi(y)$ (a). If the vortices (A, B and C) disturbed from their
original positions (dashed line) to new places (solid curve), they
will be taken away from their original positions due to pressure
difference. The stable veloctiy profile $U(y)$ has a local minimum
in vorticity $\xi(y)$ (b). The disturbed vortices (A', B' and C')
will be brought back to their original positions due to pressure
difference. } \label{Fig:stable_parallelflow_mechanism}
\end{figure}

For the two-dimensional inviscid flows with the velocity of $U$,
the vorticity $\xi=\nabla\times U$ is conserved along pathlines
\cite{Batchelor1967,Saffman1992,Huerre1998}:
 \begin{equation}
\frac{d\xi}{dt}=\frac{\partial \xi}{\partial t}+(U\cdot
\nabla)\xi=0.
 \label{Eq:stable_paralleflow_Conserve_Vor}
 \end{equation}
Its linear disturbance reduces to Rayleigh's equation provided the
basic flow $U$ being parallel. Consider an shear flow with
parallel horizontal velocity $U(y)$ in a channel, as shown in
Fig.\ref{Fig:stable_parallelflow_mechanism}. The amplitude of
disturbed flow streamfunction $\psi$, namely $\phi$, satisfies
\cite{Drazin1981,Huerre1998,CriminaleBook2003} :
 \begin{equation}
 (\phi''-k^2 \phi)-\frac{U''}{U-c}\phi=0,
 \label{Eq:stable_paralleflow_RayleighEq}
 \end{equation}
where $k$ is the nonnegative real wavenumber and $c=c_r+ic_i$ is
the complex phase speed and double prime $''$ denotes the second
derivative with respect to $y$. The real part $c_r$ is the phase
speed of wave, and $c_i\neq 0$ denotes instability. This equation
is to be solved subject to homogeneous boundary conditions
$\phi=0$ at $y=a,b$.

Consider that the velocity profile $U(y)$ has an inflection point
$y_s$ at which $U''_s=U''(y_s)=0$ and $U_s=U(y_s)$. As Sun
\cite{SunL2006a} has pointed out, the following Rayleigh's
quotient $I(f)>0$ implies the flow is stable.
\begin{equation}
I(f)=\min_{\phi} \frac{\int_{a}^{b}
[\,\|\phi'\|^2+f(y)\|\phi\|^2\,]\, dy}{\int_{a}^{b} \|\phi\|^2}
\label{Eq:stable_paralleflow_sun_Energy}
\end{equation}
where $f(y)=\frac{U''}{U-U_s}$. While if $I(f)<0$, then there is a
neutral stable model with $k_N^2=-I(f)$ and $c_r=U_s$. Moreover,
there are unstable modes with $c_r=U_s$ and $c_i\neq 0$ if
$I(f)<0$ and $0\leq k<k_N$. This can also be proved by following
the way by Tollmien \cite{Tollmien1935}, Friedrichs
\cite{Friedrichs1942,Drazin1966,Drazin1981} and Lin
\cite{LinCCBook1955}. Thus $I(f)=0$ means the flow is neutrally
stable. And there is only one neutral mode with $k=0$ and
$c_r=U_s$ in the flow. These conclusions can be summarized as a
new theorem.

Tollmien-Fridrichs-Lin theorem: The flows are neutrally stable, if
$I(f)=0$. The flows are stable and unstable for $I(f)> 0$ and
$I(f)< 0$, respectively.

From the above proof, it is obvious that the shortwaves (e.g.
$k\gg k_N$) are always more stable than longwaves (e.g. $k\ll
k_N$) in the inviscid shear flows \cite{Drazin1966,SunL2006b}. So
the shear instability is due to long-wave instability, and the
disturbances of shortwaves can be damped by the shear itself
without any viscosity \cite{SunL2006b}.

As mentioned above, the linear stability criterion can be derived
from the nonlinear one \cite{Arnold1969}. Moreover, the nonlinear
stability criterion can also be obtained from the linear one. To
illuminate this, the nonlinear criterion is retrieved
explicitly via the above theorem, %. It will be shown that the
%stability is associated with a minimum value of the total energy,
which is briefly proved as follows.

Similar to Arnol'd's definition, the general energy $H$ here is
defined as
 \begin{equation}
  H=\frac{1}{2}\int_a^b [\frac{1}{2}\|\nabla\Psi\|^2+h(\Psi)] dy
\label{Eq:stable_parallelflow_Energy_kinetic}
 \end{equation}
where $\Psi(y)$ is the streamfunction of the flow with
$\frac{\partial \Psi}{\partial y}=U(y)-U_s$, and $h(\Psi)$ is a
function of $\Psi$. The variation of $\delta H=0$ gives
 \begin{equation}
   \triangle \Psi= h'(\Psi).
\label{Eq:stable_parallelflow_Energy_varEk}
 \end{equation}
So Arnol'd's variational principle is retrieved. The function
$h=\|\nabla\Psi\|^2/2$ can also be solved from
Eq.(\ref{Eq:stable_parallelflow_Energy_varEk}). \iffalse According
to the variational principle for the inviscid flows
\cite{Arnold1965b,ArnoldBook1998}, the energy function has a
stationary value in this steady flow, on the set of all flows
isovortical to a given steady flow. by
%
 \begin{equation}
  \frac{dh}{d\Psi}=\triangle\Psi=\frac{d\Psi'}{dy}=\Psi'\frac{d\Psi'}{d\Psi}
  \end{equation}
%
Since $H$ is only associated with the velocity $U-U_s$, it is the
kinetic energy of the flow. \fi The second variation $\delta^2 H$
holds
 \begin{equation}
  \delta^2H=\frac{1}{4}\int_a^b [\|\nabla\psi\|^2+h''(\Psi)\psi^2]
  dy,
\label{Eq:stable_parallelflow_Energy_2ndvar}
 \end{equation}
where $\psi$ denotes the variation $\delta\Psi$. Noting that
$h''(\Psi)$ remains unknown, Arnol'd's nonlinear criteria have not
be extensively used. Fortunately, we can obtain the explicit
expression of $h''(\Psi)$ here via the investigation on the linear
stability criterion. For $\frac{\partial \Psi}{\partial y}=U-U_s$,
 \begin{equation}
h''(\Psi)=\frac{dh'}{d\Psi}=\frac{dh'}{dy}\frac{dy}{d\Psi}=\frac{U''}{(U-U_s)}=f(y).
\label{Eq:stable_parallelflow_confunc_2ndvar}
 \end{equation}
As $h''(\Psi)$ is solved explicitly, $\delta^2 H$ has an explicit
expression, which is greatly helpful for real applications.

Let the streamfunction of the perturbation
$\psi(x,y,t)=\phi(y)e^{i(kx-\omega t)}$, where $\omega$ is the
frequency. The averages of $\|\nabla\psi\|^2$ and $\|\psi\|^2$ are
$(\|\phi'\|^2+k^2\|\phi\|^2)/2$ and $\|\phi\|^2/2$ along the flow
direction $x$, respectively. So
Eq.(\ref{Eq:stable_parallelflow_Energy_2ndvar}) reduces to
 \begin{equation}
  \delta^2H=\frac{1}{8}\int_a^b [\|\phi'\|^2+k^2\|\phi\|^2+f(y)\|\phi\|^2]
  dy.
\label{Eq:stable_parallelflow_Energy_2ndvar2}
 \end{equation}
The sign of $\delta^2H$ is then associated with $I(f)$ in
Eq.(\ref{Eq:stable_paralleflow_sun_Energy}). If $I(f)<0$, the
second variation $\delta^2H$ can be both negative and positive,
i.e., the stationary solution is a saddle point. And $I(f)>0$
implies $\delta^2H$ is positive definite and vice versa. So the
stable flow has the minimum value of the total kinetic energy $H$.
The physical meaning of $H$ can also be revealed, as the explicit
expressions of $H$, $\frac{\partial \Psi}{\partial y}=U-U_s$ and
$h''(\Psi)=f(y)$ are obtained.

First, according to the expressions, the velocity $U$ in vorticity
conservation law Eq.(\ref{Eq:stable_paralleflow_Conserve_Vor}) can
be decomposed to two parts: the rotational flow $U-U_s$ and the
irrotational advection flow $U_s$. The vorticity $\xi$ in
Eq.(\ref{Eq:stable_paralleflow_Conserve_Vor}) depends only on
$U-U_s$, and $U_s$ is only advection velocity. Then $U-U_s$ and
$U_s$ are associated with the dynamics and kinetics of the flow,
respectively. Eq.(\ref{Eq:stable_paralleflow_Conserve_Vor})
physically shows that the vorticity field is advected by $U_s$,
which can also known from the conservation of vorticity in the
inviscid flows. A similar example is the dynamics of vortex in the
wake behind cylinder, where the vortices dominate the dynamics of
the flow and they are advected by mean flow (see Fig.2 in
\cite{Ponta2004}). The decomposition of velocity may be useful in
vortex dynamics, for that our investigation clearly shows that the
dynamics of the flow is dominated by vorticity distribution.

Then the physical meaning of $H$ can also be understood from the
above investigation. It is not $U$ but $U-U_s$ that is associated
with the general energy $H$, so $H=\frac{1}{2}\int_a^b (U-U_s)^2
dy$ is not the total kinetic energy but the kinetic energy of flow
with vorticity. Thus the stable steady states are always
minimizing the kinetic energy of the flow associated with
vorticity. This is also the reason why the flow with maximum
vorticity might be unstable, as Fj\o rtoft's criterion shows. We
would like to restate it as a theorem and to name it after Kelvin
\cite{Kelvin1875} and Arnol'd \cite{Arnold1969} for their
contributions on this field \cite{Saffman1992}.

Kelvin-Arnol'd theorem: the stable flow minimizes the kinetic
energy of flow associated with vorticity.

Both Tollmien-Fridrichs-Lin theorem and Kelvin-Arnol'd theorem are
equivalent to the following simple principle \cite{SunL2005a}: The
flow is stable, if and only if all the disturbances with $c_r=U_s$
are neutrally stable.

We have obtained the sufficient and necessary conditions for
instability, then the physical mechanism of instability can be
understood from them. The following investigation will reveal that
the essence of shear instability is due to the interaction between
the "concentrated vortex" and the corresponding resonant waves.

The concept of "concentrated vortex" comes from Fj\o rtoft's and
Sun's criteria, where the necessary conditions for instability
require that the base vorticity must be concentrated enough
\cite{SunL2006a}. We call it "concentrated vortex" for latter
convenience. The concentrated vortex is a general model of sheet
vortex in the Kelvin-Helmholtz model, for that the sheet vortex
can be recovered as the concentrated vortex
$\xi(y_s)\rightarrow\infty$ (see \cite{VallisAOFD2006} for a
comprehensive discussion about the Kelvin-Helmholtz model and
continued shear profiles). Then how the shear flow becomes
unstable, if there is a concentrated vortex? As the sufficient
condition for instability is $I(f)<0$, the normal modes in the
regime of $0\leq k< k_N$ with $c_r=U_s$ are unstable. They are
stationary or standing waves, comparing to the velocity at
inflection point. So that the shear instability is due to the
disturbance of concentrated vortex by the standing waves with
$c_r=U_s$. In this case, the resonance mechanism is valid. The
interaction waves propagate at the same speed with the
concentrated vortex, so that they are locked together and
amplified by simple advection \cite{Baines1994}. In short, the
disturbances on concentrated vortex is amplified like that in
Kelvin-Helmholtz model.

This instability mechanism combines both K-H instability and
resonance mechanism. Physically, the standing waves (with
$c_r=U_s$) can interact with the concentrated vortex, so they can
trigger instability in the flows. While the travelling waves (with
$c_r\neq U_s$) have no interaction with the concentrated vortex,
so that they can not trigger instability in the flows. This is the
mechanism of shear instability. As pointed out above, the shear
instability is due to long-wave instability. If the longwaves are
unstable, they can obtain the energy from background flows. In
this way, the energy within small scales transfer to and
concentrate on large scales. In this way, the shear instability
itself provides a mechanism to inverse energy cascade and to
maintain the large structures or coherent structures in the
complex flows.

To illuminate the mechanism of shear instability, a physical
diagram is also presented here by following the way of
interpreting the K-H instability \cite{Batchelor1967}.
Fig.\ref{Fig:stable_parallelflow_mechanism} sketches the mechanism
of shear instability in terms of wave disturbances of vortices.
The mean velocity profile $U(y)$ has an inflection point at
$y_s=0$ with $U_s=0$, and the corresponding vorticity is
$\xi(y)=-U'(y)$. There is a local maximum at $y_s$ in the unstable
vorticity profile in Fig.\ref{Fig:stable_parallelflow_mechanism}a.
According to Eq.(\ref{Eq:stable_paralleflow_Conserve_Vor}), the
vorticity is conserved in the inviscid flows. If the vortices at
the local maximum (A, B and C) are sinusoidally disturbed from
their original positions (dashed line) to new places (solid
curve), they have negative vorticities with respect to the
undisturbed ones. The vortices will induce cyclone flows around
them in consequence. The flows around the vortices become faster
(slower) in the upper (lower) of vortex A referred to the basic
flow $U(y)$. The pressures at upper and lower decrease (indicated
by - signs) and increase (indicated by + signs) according to
Bernoulli's theorem, respectively. Then vortex A gets a upward
acceleration due to the disturbed pressure difference as the
uparrow shows. This tends to take the vortex away from its
original position, so the flow is unstable. On the other hand,
Fig.\ref{Fig:stable_parallelflow_mechanism}b depicts the
disturbances in a stable velocity profile, where a local minimum
is in the vorticity profile. The disturbed vortices have positive
vorticities with respect to the undisturbed ones. The vortices
will induce anticyclone flows around them in consequence. So
vortex A' get a downward acceleration due to the disturbed
pressure difference as the downarrow shows. This tends to bring
the vortex back from its original position, so the flow is stable.
In this interpretation, the advection of $U_s$ is independent of
the shear instability, only the flow field of $U-U_s$ and the
corresponding vorticity $\xi$ are the dominations. The unstable
disturbances in Fig \ref{Fig:stable_parallelflow_mechanism} have
$c_r=U_s$, which consists with the former discussions. This
physically explains why the maximum and minimum vorticities have
different stable aspects.

\iffalse According to the theorem, there are three kinds of flows
for $f(y)<0$: (\romannumeral1) neutrally stable flows with
$I(f)=0$, (\romannumeral2) stable flows with $I(f)>0$ and
(\romannumeral3) unstable flows with $I(f)<0$. Recall the former
stability criteria, the velocity profiles can be classfied as
follows: (\romannumeral1) without inflection point,
(\romannumeral2) $f(y)>0$, (\romannumeral3) $I(f)>0$ with $f(y)<0$
and (\romannumeral4) $I(f)<0$ with $f(y)<0$. Based on the former
studies and present work, the former three categories are stable,
while the fourth category is unstable. \fi

 Though Tollmien-Fridrichs-Lin theorem is a sufficient and necessary
condition for stability, the unknown $\phi$ in the theorem
restricts its application. So some simple criteria may be more
useful. For the parallel flows within interval $a<y<b$, there are
two simple criteria.

Corollary 1: The flow is stable for $f(y)>-(\frac{\pi}{b-a})^2$
\cite{SunL2005a,SunL2006a}.

Corollary 2: The flow is unstable for $f(y)<-(\frac{\pi}{b-a})^2$
\cite{SunL2005a}.
%
\iffalse Corollary 3: The flow is unstable, if there are more than
one inflection point in the velocity profile, at which the
velocities are the same \cite{SunL2005a}.

So according to above corollaries, Tollmien's conclusion is
recovered that Fj{\o}rtoft theorem is sufficient in a symmetric or
a monotone profile.
%
Apply the theorem to unbounded flows, e.g. boundary layer flows or
free shear flows, it is obvious that the margin of instability is
$f(y)<0$ somewhere in the flows. In these cases, Fj{\o}toft's
theorem is also sufficient for instability. As Tollmien
\cite{Tollmien1935} proved the same result for symmetrical channel
flows and monotone profiles. Here we point out that Tollmien's
result can be extended to any unbounded shear flows.
%
\fi

\iffalse Similar to Kuo's criterion \cite{KuoHL1949}, the criteria
for stability of parallel inviscid flow can be applied to
barotropic geophysical flow \cite{SunL2006a}. This extension is
trivial for the cases of $f$-plane and $\beta$-plane, and is
omitted here. \fi

In summary, the general stability and instability criteria are
obtained for inviscid parallel flow. And the criterion is
associated with a minimum value of energy, which shows the
relationship between linear and nonlinear stability criteria. Then
the mechanism of shear instability is investigated, which is
explained as the resonance of standing waves with the concentrated
vortex at $c_r=U_s$. A physical explanation is also sketched.
Finally, some useful criteria are given. In general, these
criteria will lead future works to investigate other instabilities
in hydrodynamics.

 The work was original from author's dream of
understanding the mechanism of instability in the year 2000, when
the author was a graduated student and learned the course of
hydrodynamics stability by Prof. Yin X-Y at USTC.

\iffalse

\fi
%The work is purely due to personal interesting, and it does not be
%supported by any official founds.

%The author thanks Prof. Yin X-Y and Prof. Sun D-J  for simulation
%discussions.

\iffalse

\bibliography{MSH1}

\fi

\end{document}